\documentclass[twocolumn,showpacs,amsmath,amssymb,prl]{revtex4}

\usepackage{epsfig}

\newcommand{\be}{\begin{equation}}
\newcommand{\ee}{\end{equation}}
\newcommand{\ba}{\begin{eqnarray}}
\newcommand{\ea}{\end{eqnarray}}
\begin{document}
\title{ Anisotropic Fermi superfluid via  p-wave Feshbach resonance }
\author{Chi-Ho Cheng and S.-K. Yip}
\affiliation{ Institute of Physics, Academia Sinica, Nankang,
Taipei 115, Taiwan}
\date{\today}

\begin{abstract}
We investigate theoretically Fermionic superfluidity induced by
Feshbach resonance in the orbital p-wave
channel.  We show that, due to the dipole interaction, the pairing
is extremely anisotropic.  When this dipole interaction is
relatively strong, the pairing has symmetry $k_z$. When it is
relatively weak, it is of symmetry $k_z + i \beta k_y$
(up to a rotation about $\hat z$, here $\beta < 1$).
 A phase transition between these two states can
occur under a change in the magnetic field or the density of the gas.
\end{abstract}

\maketitle

In the earlier investigations of trapped Bose and Fermi gases,
the interactions between particles are in general weak
since the gases are typically dilute, with the interparticle
distances much larger than the scattering $a$ between the
particles (see, e.g., \cite{Dalfovo99,Castin01} and reference
therein).   Recently however, it has been recognized
that the interaction among particles can be tuned with
the use of Feshbach resonances \cite{Tiesinga92}.  In the case
of two Fermion species and a Feshbach resonance between
them, one can tune, by varying the magnetic field and
hence the resonance level energy, the interaction from
(its "background" value through) weakly to strongly attractive.
The ground state of the system is expected to evolve
from a Bardeen-Cooper-Schrieffer (BCS) superfluid
with long-range (compared with interparticle distances)
Cooper pairing to a Bose-Einstein condensate (BEC) of
tightly bound molecules.  \cite{cross0}
This cross-over has already been a subject of
intensive and experimental investigations.
\cite{crosst,crosse}

Practically all the above investigations are for the
case where the resonance involves an s-wave bound
state in the "closed" channel.  Thus the aforementioned
states involve pairing only in the s-wave state.  More
recently, it has been demonstrated experimentally
also that p-wave Feshbach resonances exist \cite{Regal03,Ticknor04,Zhang04}.
This thus raises the possibility of superfluids with p-wave
Cooper pairing or BEC of p-wave molecules \cite{Ho}.

Consider now therefore a single Fermion species with an effective
interaction induced by the p-wave Feshbach resonance. This brings
in mind the well-known system of superfluid $^3$He.
\cite{Leggett75} $^3$He has (nuclear) spin $1/2$, in general {\it
not} spin-polarized and basically spatially isotropic.  Back in
the 60's, Anderson and Morel \cite{Anderson61} investigated
theoretically the superfluid state for this system by assuming
that the pairing exists only between the same (say ${\uparrow}$)
species.  They found that the ground state of this system
corresponds to Cooper pairing in the $l = 1$, $m = 1$ channel,
i.e., the pair-wavefunction has the symmetry of the spherical
harmonics $Y_1^1(\hat k) \propto (\hat k_x + i \hat k_y)$
 (or its spatial rotation). This state is
realized in the $^3$He A-phase and is known as the
"axial" phase \cite{Leggett75}. One can thus conclude immediately
that for a spin-polarized but otherwise spatially isotropic
system, the pairing is again expected to be in the $Y_1^1 (\hat
k)$ state.  This argument is in agreement with the findings
of Ref \cite{Ho}.

 However, in our atomic system of interest, the
interaction is far from isotropic.  In particular, as demonstrated
and explained by Ticknor et al \cite{Ticknor04} 
for the case of $^{40}$K, due to the
magnetic dipole of the alkali atoms, the $l=1$, $m=0$ resonance
occurs at a higher magnetic field than the $m = {\pm} 1$ ones.
For a given magnetic field, the induced effective 
interaction is actually more attractive in the $m = 0$ channel
than the $m = {\pm} 1$ ones. It is now thus non-trivial what the
symmetry of ground state should be.  This is the specific question
on which we would concentrate in this paper.

\begin{figure}[tbh]
\begin{center}
\includegraphics[width=3in]{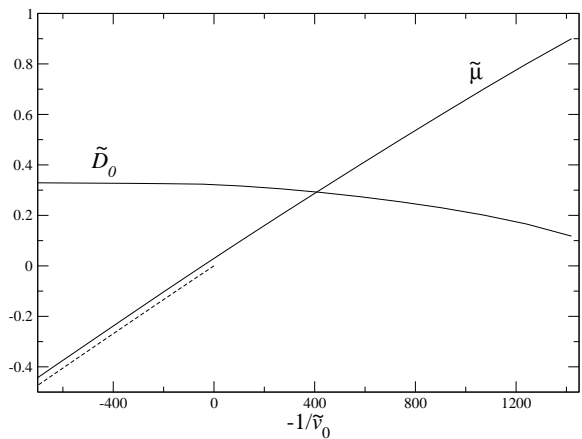}
\end{center}
 \caption{The dimensionless parameters $\tilde D_0$ and $\tilde
 \mu$ as functions of $-1/\tilde v_0$. 
$\tilde c_0= \tilde c_1=-100$ and $\tilde D_{\pm 1} =0$ in 
this case. The dashed line represents $ - \epsilon_b / 2 \epsilon_F$.}
 \label{gap_c100}
 \vspace{-5pt}
\end{figure}

\begin{figure}[tbh]
\begin{center}
\includegraphics[width=3in]{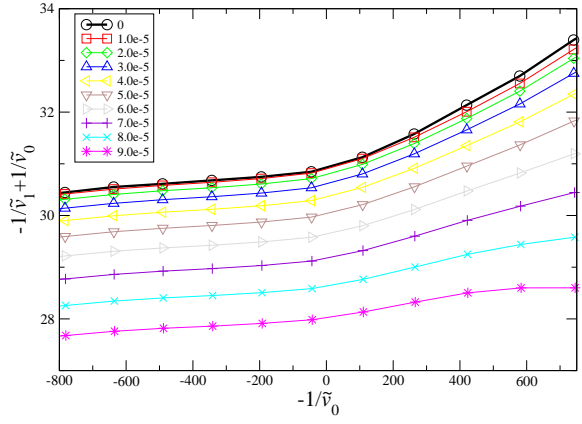}
\end{center}
 \caption{Contour plot of $\tilde D_{\pm 1}$ as a function of
$-1/\tilde v_1+ 1/\tilde v_0$ and $-1/\tilde v_0$ for
 $\tilde c_0= \tilde c_1= -100$. The line for $\tilde 
D_{\pm 1} =0$ corresponds to the critical value $-1/\tilde v_1^*+ 1/\tilde v_0$.}
 \label{mixed2}
 \vspace{-5pt}
\end{figure}

Our findings are basically as follows.  For a sufficiently dilute
Fermi gas or for the case where the $m = 0$ and $m = {\pm} 1$
resonances are sufficiently far apart, the pairing occurs only in
the $m=0$ channel.  That is,  the ground state is a BCS state with
Cooper pairing of symmetry $Y_1^0 (\hat k) \propto k_z$ for
"large" magnetic fields and a BEC state of $m=0$ Bosons for
magnetic fields sufficiently below the resonance(s).    There is
no pairing or Boson in the $m = \pm 1$ channel.  
The orbital symmetry of this pairing is the same as the "polar"
phase in the $^3$He literature \cite{Leggett75}. For a
sufficiently dense gas or for the case where the resonances are
sufficiently closed to each other, the BCS case at high magnetic
field corresponds to a state with Cooper pairing $\propto 
\left[ (\hat k
\cdot \hat z) + i \beta (\hat k \cdot \hat a) \right]$.
  Here $\hat z$ is
along the magnetic field direction and $\hat a$ is a vector in the
plane perpendicular to $z$, and $\beta$ is a number less than
unity and is field dependent. 
This state is thus intermediate between the polar
and axial phases. At lower fields, $\beta$ may vanish
and again the BCS pairing or the BEC condensation is again
entirely in the $m=0$ channel.  These results are summarized in
Figs. \ref{gap_c100} and \ref{mixed2}.

We begin with the Hamiltonian $H = H_f + H_b + H_{\alpha}$ where
\be
H_f = \sum_{\vec k} (\frac{\hbar^2 k^2}{2M} - \mu)
 a_{\vec k}^{\dagger} a_{\vec k}
\label{Hf}
\ee
\be
H_b = \sum_{\vec q} \left( - 2 \mu + \delta_m +
    \frac{\hbar^2 q^2}{4M} \right)
   b_{\vec q, m}^{\dagger} b_{\vec q, m}
\label{Hb}
\ee
 \begin{widetext}
\be
H_{\alpha} = \frac{1}{L^{3/2}}
   \sum_{m, \vec q, \vec k}
  \left\{ b_{\vec q, m}^{\dagger}
   \left( i \sqrt{ 4 \pi} k Y_1^{m *} (\hat k) \tilde \alpha_m^{*} \right)
     a_{- \vec k + \vec q/2} a_{\vec k + \vec q/2}
   + {\it h.c.}   \right\}
\label{Hal}
\ee
 \end{widetext}

This Hamiltonian is a generalization of the ones already
commonly employed \cite{crosst,Fesh,renorm,Diener04} for
s-wave Feshbach resonances to the case of several
$l =1$, $m = 0, \pm 1$ closed channels.
$H_f$ and $H_b$ are the Hamiltonians for the free Fermions and
Bosons (particle pairs in the "closed" channels)
 respectively, and $H_{\alpha}$ represents
the Feshbach coupling.
$a_{\vec k}$ is the annihilation operator for a Fermion with
momentum $\vec k$,  $b_{\vec q}$ the corresponding
operator for a Boson with angular momentum $l = 1$ and
$z$-axis projection $m$ with momentum $\vec q$,
$\mu$ and $M$ are the chemical potential and mass of
the Fermions.  In $H_{\alpha}$,
the factor $Y_1^m (\hat k)$ reflects the symmetry of
the $l, m$ bound state and the linear factor in $k$ arises
from the small momentum approximation for the coupling.
$\delta_m$ is the (bare) detuning of the energy of
the closed channel with angular momentum projection $m$,
and $\tilde \alpha_m$ the corresponding coupling constant.
For low energy scattering between a pair of particles
of momentum $\pm \vec k$, the scattering amplitude
in the $l=1$, $m$ partial wave can be parameterized
as, using the same notations as ref \cite{Ticknor04},
\be
f_m(k) =  \frac{k^2}{ - \frac{1}{v_m} + c_m k^2 - i k^3}
\label{fm}
\ee
$v_m$ has the dimension of a volume and $c_m$ an inverse length.
The magnetic field dependent parameters $v_m$, $c_m$ are in
principle available experimentally and have
already been measured for $^{40}$K for atoms
in the $ |f, m_f> = | 9/2, -7/2>$ hyperfine states. \cite{Ticknor04}
The bare parameters $\delta_m$ and $\alpha_m$ can be related to the physical
parameters $v_m$, $c_m$ by considering scattering
between two Fermions using the Hamiltonian given in
eqs (\ref{Hf}-\ref{Hal}). \cite{Ho}
These relations are:
\be
- \frac{1}{v_m} =
\frac{ 4 \pi \hbar^2}{M} \left(
  \frac{\delta_m}{|\tilde \alpha_m|^2}
   -  \frac{1}{L^3} \sum_{\vec k} \frac{M}{\hbar^2} \right)
\label{vm}
\ee
\be
c_m = - \frac{ 4 \pi \hbar^2}{M} \left(
  \frac{\hbar^2}{ M |\tilde \alpha_m|^2}
   +  \frac{1}{L^3} \sum_{\vec k} \frac{1}{2 \epsilon_k} \right)
\label{cm}
\ee
Here $\epsilon_k \equiv \frac{k^2}{2M}$.
The divergent sums over $\vec k$ on the right-hand-sides of the above two
equations can be regulated either by introducing a cut-off
or invoking the fact that the coupling $\tilde \alpha_m$ must
actually decay to zero at large momenta.
Below we shall express all physical quantities in $v_m$ and $c_m$
in the final expressions and omit these explicit cutoffs.

With Feshbach resonance for the sub-channel $m$, $1/v_m$ is field
dependent, vanishing at a field $B_m^{*}$.  In contrast,
$c_m$ has a definite sign. For the ease of discussions,
we shall assume that $c_m < 0$ and field independent,
$-1/v_m$  is an increasing function of field
  $- 1/v_m > (<) 0$
for $B > (<) B_m^{*}$, as in the case of $^{40}$K.
(This corresponds to the case where $\delta_m$ is an increasing
function of field and $\tilde \alpha_m$ weakly field
dependent, c.f. eqs (\ref{vm}) and (\ref{cm})).
For $B < B_m^{*}$, a bound state appears.  The energy
of this bound state is given by 
$- \epsilon_{b,m} = - \hbar^2 {\kappa_m}^2/M$
with $k = i \kappa_m$ being a pole for $f_m (k)$.
For small detuning below the resonance,
$\kappa_m^2 = 1/ [ (-c_m)(v_m) ]$.  Since $1/v_m$
should be roughly linear in $B$ near the resonance,
$\epsilon_{b,m}$ increases linearly with $ (B_m^{*} - B)$
(in contrast to s-wave, where it is quadratic).

Moreover, as explained in \cite{Ticknor04}, due to
the dipole interaction, $B_0^{*} > B_{1}^{*} = B_{-1}^{*}$.
Thus in the field range of interest, $ - 1/v_1 = -1/v_{-1} > -1/v_0$.
We can say that, at a given field, the effective interaction
between the Fermions is less attractive for relative
angular momentum projections $m = \pm 1$ than $m = 0$.

Now we proceed to find the ground state for the many-body problem.
We assume a mean-field theory and replace $b_{\vec q, m}$ by
c-numbers.  Only its $\vec q =0$ value is non-vanishing. It is
convenient to introduce the symbols
 \be D_m = - i \sqrt{ 4 \pi}
\tilde \alpha_m b_{0,m} / L^{3/2} \label{defDm} 
\ee
 \be
\Delta_{\vec k} = \sum_m D_m k Y_1^m (\hat k) 
\label{defDelta} 
\ee
so that $H_{\alpha}$ becomes $\Delta^{*}_{\vec k} a_{-\vec k}
a_{\vec k}
   + {\it h.c.} $.  The Fermionic part of the Hamiltonian
can be solved by Bogoliubov transformation.
The value for $b_{0,m}$ can also be easily found since
the Hamiltonian is quadratic in this variable.
 In terms of $D_m$, we get
\be
( -2 \mu + \delta_m ) D_m
  = \frac { 4 \pi} {L^3}  \sum_k   |\tilde \alpha_m|^2
    k   Y_1^{m *}(\hat k)
   \frac {\Delta_{\vec k}} { 2 \left[ (\epsilon_k-\mu)^2+|\Delta_{\vec k}|^2 
\right]^{1/2}  }
\label{Dm} 
\ee 
Using eq (\ref{vm}) and (\ref{cm}) we obtain
 \begin{widetext}
\be - \frac{M}{4\pi v_{0}} D_{0} + \frac{M^2 c_0}{2\pi} \left\{ D_0
 \left[\mu - \frac{3M}{20\pi}(2|D_1|^2+3|D_0|^2+2|D_{-1}|^2) \right]
    + \frac{3M}{10\pi}D_{0}^*D_1 D_{-1}
\right\}
 = \frac{2\pi}{L^3}\sum_{\vec k, m}D_m Y_1^{0 *}(\hat k) Y_1^{m}(\hat
 k) h(\vec k)
 \label{D0} \ee 

\be - \frac{M}{4 \pi v_{1}} D_{1} + \frac{M^2 c_1}{2\pi}
 \left\{ D_1 \left[\mu - 
\frac{3M}{20\pi} (3|D_1|^2+2|D_0|^2+6|D_{-1}|^2) \right] +\frac{3M}{20\pi}D_{-1}^*D_0^2 
\right\}
 = \frac{2\pi}{L^3}\sum_{\vec k, m}D_m Y_1^{1 *}(\hat k) Y_1^{m}(\hat
 k) h(\vec k)
 \label{D1} \ee
\end{widetext}
and a corresponding equation with $m= 1 \leftrightarrow m=-1$.
Here
 \be
 h(\vec k) \equiv \frac{k^2}{\left[ (\epsilon_k-\mu)^2 + |\Delta_{\vec k}|^2 
\right]^{1/2} }
 - \frac{k^2}{\epsilon_k}(1+\frac{\mu}{\epsilon_k}-\frac{|\Delta_{\vec 
k}|^2}{2\epsilon_k^2})  \ .
 \ee
 These equations are to be solved
together with the number equation 
\be n = \frac{1}{L^3} \left(
\sum_{\vec k} \langle a_{\vec k}^{\dagger} a_{\vec k} \rangle
   + 2 \sum_m |b_{0,m}|^2 \right)
\label{n1} \ee 
 $ \langle a_{\vec k}^{\dagger} a_{\vec k} \rangle$
is given by 
 $ v_{\vec k}^2 \equiv \frac{1}{2}(1-\frac{\epsilon_k - \mu}{[
(\epsilon_k-\mu)^2+|\Delta_{\vec k}|^2]^{1/2}})$. Since $\Delta_{\vec
k}$ from eq (\ref{defDelta}) is linear in $k$, the sum in eq (\ref{n1})
over $\vec k$ is formally divergent due to the large $\vec k$
contributions. However, we can regularize it by employing again eq
(\ref{cm}). We have finally \be n = \frac{1}{L^3} \sum_{\vec k}
\left( v_{\vec k}^2
           - \frac{ |\Delta_{\vec k} |^2} { 4 \epsilon_k^2 } \right)
   + \frac{M^2}{(4\pi\hbar)^2}\sum_m(-c_m)|D_m|^2
\label{n}
\ee

Eq (\ref{D0}), (\ref{D1}), (\ref{n}) are our principal equations,
with parameters characterizing the Feshbach resonances expressed
entirely in $v_m$ and $c_m$. These equations determine the order
parameters $D_m$ and chemical potential $\mu$ for given density
$n$ and "interaction parameters" $v_m$ and $c_m$. For simplicity,
in writing these equations we have already dropped the terms with
explicit $1 / |\tilde \alpha_m|^2$ factors.  
These terms are small under the 
"wide-resonance" regime \cite{crosst}(b), \cite{renorm,Diener04}.

Since the interaction is less attractive for angular momentum
projections $m = \pm 1$, for sufficiently large difference between
$-1/v_{0}$ and $-1/v_{\pm 1}$ we expect (and verify below) that
the pairing is entirely in the $m = 0$ partial wave.  We thus
first begin our analysis by assuming that only $D_0$ is
non-vanishing.  eqs (\ref{D0}) and (\ref{n}) can be solved
simultaneously similar to the s-wave case. It is convenient to
express the results in dimensionless form. 
We define $\tilde \mu\equiv
\mu/\epsilon_{\rm F}$, $\tilde D_m \equiv D_m/v_{\rm F}$, $\tilde
c_m \equiv n^{-1/3}c_m$, and $\tilde v_m \equiv n v_m$ where $\epsilon_{\rm
F} \equiv \hbar^2 k^2_{\rm F}/2M$, $v_{\rm F} \equiv k_{\rm F}/M$, 
and $k_{\rm F} \equiv 6\pi^2 n$. The results are as shown in 
Fig. \ref{gap_c100}  (for the case
$\tilde c_0 = \tilde c_1 = - 100$, see below for 
the reason of this choice). 
In the BCS regime ( large $-1/\tilde v_0 \gg 1 $ or $B -
B^*_0$, not shown explicitly), $\tilde \mu \to 1$, and $\tilde D_0$,
being proportional to the magnitude of the BCS gap,
is $\ll 1$.
In the BEC ( $ -1/\tilde v_0 \ll  -1 $ or large and negative $ B - B^*_0$), 
$\mu$ is approximately $- \epsilon_b / 2 $
and $\tilde D_0$ approaches a constant.  This latter value can be obtained
from eq (\ref{n}) as $ (32 \pi/3)^{1/3} (- \tilde c_0)^{-1/2} $.

The "cross-over" behavior in Fig \ref{gap_c100}
is analogous to the s-wave case, where
the corresponding x-axis is 
$ x = - 1/ (n^{1/3} a)$ where $a$ is the s-wave scattering length.
 Note here
$D_0$ has the dimension of (energy $\times$ inverse length) and  
behaves differently from the s-wave $\Delta$
in the BEC limit.  
We have also performed calculations for other values of $\tilde c_0$.  
The size of the crossover region is roughly proportional to the
value of $\tilde c_0$.  For example, for $\tilde c_0 = -200$,
the corresponding results can be captured well by replacing the
x-axis by $ - 1/ 2 \tilde v_0$ and dividing $\tilde D_0$ by 
$1 / \sqrt{2}$ in Fig \ref{gap_c100}.
The result in Fig \ref{gap_c100} is similar to that 
 in Ref \cite{Ho}, even though the latter
actually studied a different ($D_1 \ne 0$, $D_0 = D_{-1} = 0$) state.

The above behavior applies only to sufficiently large
$ -1/v_{\pm 1} - ( -1/v_0) > 0$.  When this difference
is sufficiently small, $D_{\pm 1}$ will become finite.
The critical value for $1/v_{\pm 1}$,
denoted by $1/v_{\pm 1}^{*}$, can be found
by putting $D_{\pm 1} = 0$ in eq (\ref{D0}) and
linearizing eq (\ref{D1}) in $D_{\pm 1}$.
We obtain, for $c_1=c_0$,
\begin{eqnarray}
&&-\frac{1}{\tilde v_1^*}+\frac{1}{\tilde v_0} =
\frac{3(6\pi)^\frac{2}{3}}{5\pi}{\tilde D_0}^2(-\tilde c_0)
\nonumber \\
&& + 9\pi \int_0^\infty dx \int_{-1}^1 dy
 \frac{x^4(1-3 y^2)}{[(x^2-\tilde \mu)^2+\frac{3}{\pi}\tilde D_0^2 x^2 
y^2]^{1/2}}
\label{v*}
\end{eqnarray}
In the BCS limit ($-1/\tilde v_0 \gg 1$), 
the first term in eq (\ref{v*}) is negligible whereas
the second term becomes a constant independent of $D_0$.
From this we get
$-1/\tilde v_1^* +
1/\tilde v_0 \rightarrow 12\pi \approx 37.7$.
In BEC limit
($-1/\tilde v_0 \ll -1$),
the main contribution comes from the first
term in eq (\ref{v*}).  Using the aforementioned
asymptotic values of $\tilde D_0$ we get
 $-1/\tilde v_1^* + 1/\tilde v_0
\rightarrow 48 \pi /5 \approx 30.2$.
$-1/\tilde v_1^* + 1/\tilde v_0$ is shown as the
 thick black line ($\tilde D_0 = 0$)
in Fig \ref{mixed2}.

For $-1/\tilde v_1 + 1/\tilde v_0 $ less than this 
critical value, $D_{\pm 1}$ are finite.
Eqs (\ref{D0}), (\ref{D1}) involve
three complex unknowns $D_m$.  By gauge invariance
we can always choose $D_0$ to be real.  Under this choice,
the solutions we found belong to the class $D_{1} = D_{-1}^{*}$.  
Writing $D_{1} = |D_{1}| e^{i \chi}$,
$\Delta_{\vec k}$ then has
the angular dependence $\propto$
$ D_0 \hat k_z + i \sqrt{2} |D_1| \hat k \cdot \hat a$
$\propto (\hat k_z + i \beta \hat k \cdot \hat a)$
where $\hat a = ({\rm cos} \chi) \hat y + ({\rm sin} \chi) \hat x$
is a unit vector perpendicular to $\hat z$ and
$\beta = \sqrt{2} |D_{1} / D_{0}|$.
A particular solution is given by the case where
$D_1$ and $D_{-1}$ are both real where $\hat a = \hat y$.
The other solutions are simply related to this one
by a rotation about $\hat z$.  Without loss of
generality we shall therefore pretend that
$D_{m}$'s are all real.

The contours of the order parameters 
$D_{\pm 1}$ are also shown in Fig \ref{mixed2}.
In the $D_{\pm 1} = 0$ phase, the state is rotationally
invariant about $\hat z$, whereas this symmetry
is broken in the $D_{\pm 1} \ne 0$ phase. 
There is a (quantum) phase transition between these
two phases when one
crosses the line $-1/\tilde v_1^* + 1/\tilde v_0$.

For the $^{40}$K case studied in \cite{Ticknor04},
the Feshbach resonances are at
$B^*_0 \approx 198.8 G$ and $B^*_1 \approx 198.4G$.  There,
$c_1$ and $c_0$ are both only weakly field dependent
and are approximately given by $-0.02 a_0^{-1}$.
Our choice of $\tilde c = - 100$ above corresponds
to a density of roughly $6.7 \times 10^{13} {\rm cm}^{-3}$.
Near the resonant fields, $ -1 / v_1 + 1 / v_0 \approx 
2.1 \times 10^{-8} a_0^{-3}$ and is roughly
field independent.   
Thus the density determines the values for both
$\tilde c_{0, 1}$ and $ -1 / \tilde v_1 + 1 / \tilde v_0$
while varying the magnetic field corresponds roughly 
to moving along a horizontal line on our
phase diagram of Fig. \ref{mixed2}
(with increasing field towards the right and
the distance of the line from the x-axis 
proportional to $n^{-1}$).

While preparing this manuscript, we become aware of
Ref \cite{Rad05} which studies essentially the same
problem.   Whereas our results agree with \cite{Rad05} for
very large and very small splitting between the resonances,
the conclusions differ in the intermediate splitting regime.
Our prediction is that the state is
 $\sim \hat k_z + i \beta \hat k_y$ on the BCS
side whereas it should be $~ \hat k_z$ on the BEC side.
  Their conclusion is the opposite.
 The reason for this disagreement is not yet understood.
We believe that our results are more reasonable.
For large positive detuning, the splitting should be less relevant
and the pairing state should resemble more that of the isotropic system.
 On the BEC side, the system should be closer to  
a Bose condensate of lowest energy ($\hat k_z$) molecules. 

In conclusion, we have shown that p-wave Feshbach resonance
in general leads to anisotropic Fermi superfluids.  The
symmetry of the ground state depends on both the density
and magnetic fields.

One of us (SKY) is very grateful to T.-L. Ho and A. J. Leggett for
educating him about how to treat theoretically Feshbach resonances
in many-Fermion systems during his attendance of the workshop
"Quantum Gases" held at the Kavli Institute for Theoretical
Physics, University of California, Santa Barbara. This research
was supported by the National Science Foundation under grant
number PHY99-07949 (SKY), NSC93-2112-M-001-016, and
NSC93-2816-M-001-0007-6.


\vspace{15pt}


\vspace{15pt}


\end{document}